\documentclass[runningheads]{llncs}
\usepackage[T1]{fontenc}
\usepackage{amsmath}
\usepackage{amssymb}
\usepackage{graphicx,verbatim}
\usepackage{xcolor} 
\newcommand{\best}[1]{\textcolor{red}{{#1}}}
\newcommand{\second}[1]{\textcolor{blue}{#1}}
\usepackage{tabularx, multirow}
\begin{document}
\title{Energy-Mamba: A Physics-Constrained State-Space Model for Medical Image Classification}

\author{Mohamed Mabrok\inst{1}\orcidID{0000-0003-3638-4424}\and Yalda Zafari\inst{1}\orcidID{0009-0004-9737-7134}\and Essam A. Rashed\inst{2, 3}\orcidID{0000-0001-6571-9807}} 
\authorrunning{M. Mabrok et al.}
\institute{Department of Mathematics and Statistics, Qatar University, Doha, Qatar \and Graduate School of Information Science, University of Hyogo, Kobe 650-0047, Japan\and Advanced Medical Engineering Research Institute, University of Hyogo, Himeji 670-0836, Japan}

\maketitle             
\begin{abstract}
State-Space Models (SSMs), particularly Mamba, offer linear-time complexity for long-range dependencies, making them attractive for medical imaging with limited annotated data. However, adapting these sequential models to 2D images through unconstrained state evolution causes representational drift, the dynamic hidden state progressively loses fidelity to local image features. We introduce Energy-Mamba, integrating SSM dynamics with physics-informed constraints via a learnable potential energy function that quantifies compatibility between evolving states and static local features. Our Energy-Mamba Block introduces a gradient-based forcing term, computed dynamically via automatic differentiation, that pulls states toward low-energy configurations maintaining local visual fidelity. This formulation mirrors Hamiltonian dynamics: kinetic energy (SSM scan) plus potential energy (our constraint function) govern state trajectories. This architectural prior enables learning implicit constraints for robust, faithful representations, crucial in medical imaging where fine-grained local detail drives accurate diagnosis. Evaluated on four datasets (retinal OCT, chest X-ray, microscopy, abdominal CT), Energy-Mamba achieves state-of-the-art classification performance with significantly fewer parameters, demonstrating that physics-informed grounding can enhance both efficiency and representational quality in medical vision tasks.


\end{abstract}
\section{Introduction}

Medical image analysis demands models capable of capturing both fine-grained local detail and holistic global context across diverse imaging modalities. Local features, such as lesion boundaries, tissue textures, and cellular morphology, are essential for precise localization and characterization, while global understanding enables the model to integrate spatial relationships, anatomical context, and inter-regional dependencies that are often equally critical for accurate diagnosis~\cite{zafari2024transformers}. Deep learning has driven remarkable progress in this domain, yet simultaneously modeling local and global representations with sufficient capacity remains a persistent challenge, compounded further by the data-scarce conditions typical of clinical settings, where large annotated datasets are costly and often unavailable.

Convolutional neural networks (CNNs) excel at extracting local features but struggle to model long-range spatial dependencies without deep stacking or large receptive fields~\cite{sarvamangala2022convolutional}. Vision Transformers (ViTs)~\cite{dosovitskiy2020image} address this through global self-attention, achieving strong performance on medical imaging benchmarks, but their quadratic complexity with respect to sequence length renders them prohibitively expensive for high-resolution volumetric data. These limitations have motivated the exploration of State-Space Models (SSMs) as an efficient alternative~\cite{liu2024vmamba}.

Mamba~\cite{gu2023mamba}, built upon selective SSMs, achieves linear-time complexity while modeling long-range dependencies, making it an attractive backbone for medical image analysis. Recent vision adaptations such as VMamba~\cite{liu2024vmamba} extend Mamba to 2D spatial data via multi-directional selective scanning (SS2D). Despite their efficiency, these architectures suffer from a fundamental limitation: their state evolution is purely \textit{kinetic}. As tokens sequentially modulate the SSM parameters, the hidden state undergoes continuous transformation with no mechanism to anchor it to local image content. This \textit{representational drift}~\cite{mabrok2026controllability,hardan2025flatten} causes earlier spatial information to progressively lose fidelity across layers, a particularly critical failure mode in medical imaging where local detail, the boundary of a lesion, the texture of a nodule, is directly diagnostic.

Existing efforts to adapt SSMs for medical vision have largely focused on hybrid architectures that combine CNN and SSM blocks to leverage local inductive biases alongside long-range sequential modeling. These designs have demonstrated strong performance across tasks including segmentation~\cite{kui2025gl,xu2025mambavesselnet++} and classification~\cite{zafari2025hybrid,yue2024medmamba}. However, while some studies have observed instability in hidden state representations under sequential scanning~\cite{mabrok2026controllability,hardan2025flatten}, no prior work has directly addressed the underlying cause: the absence of a restorative constraint on the hidden state trajectory. Existing hybrid designs mitigate symptoms of drift through architectural diversity, but do not prevent it, leaving the fundamental vulnerability of unconstrained kinetic state evolution unresolved.

We take a different approach, drawing inspiration from classical mechanics. In physical systems, a potential energy field provides a restoring force that continuously guides a system's trajectory toward stable, low-energy configurations. We propose \textbf{Energy-Mamba}, which introduces this principle directly into the SSM framework through a learnable potential energy function. This function quantifies the compatibility between the evolving hidden state and the static patch embedding at each spatial location. Its negative gradient defines a physics-inspired forcing term, computed via automatic differentiation, that continuously corrects the state trajectory to maintain fidelity to local visual content. The resulting block admits a clean Hamiltonian interpretation: the SSM scan governs kinetic evolution while the energy function governs potential constraints, and together they define a stable, interpretable dynamical system. Our contributions are as follows:
\begin{itemize}
    \item We identify representational drift as an inherent limitation of vision Mamba architectures and provide a physics-informed theoretical framework to address it.
    \item We propose the Energy-Mamba block, which augments SSM dynamics with a learnable potential energy function and a gradient-based forcing term grounded in Hamiltonian mechanics, introducing an explicit restorative constraint on hidden state evolution.
    \item We demonstrate competitive or state-of-the-art classification performance on four MedMNIST benchmarks spanning diverse imaging modalities, while using substantially fewer parameters (2.0M) than all compared methods.
\end{itemize}

\section{Methods}
\label{sec:method}

\subsection{Mamba-Based Architectures}

The Mamba architecture~\cite{gu2023mamba} is grounded in continuous-time state-space models (SSMs). For vision tasks, it operates as a multi-input-multi-output (MIMO) system parallelized across $D$ channels. For a $D$-channel input $\mathbf{x}(t) \in \mathbb{R}^D$, the system maintains a latent state $\mathbf{h}(t) \in \mathbb{R}^{D \times N}$, where $N$ is the state dimension per channel. Each channel $d$ evolves according to:
\begin{equation}
    \dot{\mathbf{h}}_d(t) = \mathbf{A}_d \mathbf{h}_d(t) + \mathbf{B}_d x_d(t), \quad
    y_d(t) = \mathbf{C}_d \mathbf{h}_d(t) + D_d x_d(t)
    \label{eq:ssm_cont}
\end{equation}
where $\mathbf{A}_d \in \mathbb{R}^{N \times N}$ is the state matrix, and $\mathbf{B}_d$, $\mathbf{C}_d$, $D_d$ are channel-specific projection parameters. This continuous system is discretized using a timescale parameter $\Delta$ via zero-order hold, yielding:
\begin{equation}
    \bar{\mathbf{A}}_d = \exp(\Delta_d \mathbf{A}_d), \quad
    \bar{\mathbf{B}}_d = (\Delta_d \mathbf{A}_d)^{-1}(\exp(\Delta_d \mathbf{A}_d) - \mathbf{I}) \cdot \Delta_d \mathbf{B}_d
    \label{eq:discretization}
\end{equation}
The resulting discrete recurrence is $\mathbf{h}_{k,d} = \bar{\mathbf{A}}_d \mathbf{h}_{k-1,d} + \bar{\mathbf{B}}_d x_{k,d}$. Mamba's core innovation (S6)~\cite{gu2023mamba} makes the parameters $\Delta$, $\mathbf{B}$, and $\mathbf{C}$ input-dependent, enabling selective attention over the sequence. Stability is enforced by constraining $\mathbf{A} = -\exp(\mathbf{A}_{\mathrm{logs}})$, ensuring all eigenvalues are negative.

For vision tasks, the 2D Selective Scan (SS2D) module~\cite{liu2024vmamba} extends this mechanism to spatial data by scanning feature maps along multiple traversal paths. However, this purely kinetic state evolution introduces a critical challenge: as tokens sequentially modulate the SSM parameters, earlier spatial information tends to drift toward different representational subspaces. Scan direction strongly governs which regions are amplified or suppressed, and small input perturbations can accumulate into substantial representational changes across layers, a phenomenon termed \textit{representational drift}~\cite{mabrok2026controllability}. Since the hidden state is updated incrementally rather than anchored to a fixed reference, the model lacks an intrinsic restoring mechanism.

\subsection{Proposed Approach: Energy-Mamba}

To address representational drift, we propose the \textbf{Energy-Mamba} block, which augments the standard Mamba block with a physics-inspired potential energy mechanism. Each block receives two inputs: (1) the \textit{dynamic state} $\mathbf{Z} \in \mathbb{R}^{B \times H' \times W' \times C}$, the evolving feature map from the previous block; and (2) the \textit{static features} $\mathbf{\Phi} \in \mathbb{R}^{B \times H' \times W' \times C}$, the patch embedding from the initial layer, which serves as a fixed spatial reference throughout the network.

\noindent\textbf{Potential Energy Function.}
We define a learnable scalar \textit{potential energy function} $\mathcal{E}: \mathbb{R}^C \times \mathbb{R}^C \rightarrow \mathbb{R}$ that quantifies the compatibility between the dynamic state and the static features at each spatial location $(i,j)$. The dynamic state is first normalized: $\mathbf{H}_{\mathrm{norm}} = \mathrm{LayerNorm}(\mathbf{Z})$. The energy at each position is then computed by a lightweight MLP applied to the concatenation of the two feature vectors:
\begin{equation}
    \mathbf{z}_{i,j} = \mathrm{concat}(\mathbf{h}_{i,j},\, \boldsymbol{\phi}_{i,j}) \in \mathbb{R}^{2C}, \quad
    E_{i,j} = f_{\mathrm{energy}}(\mathbf{z}_{i,j})
    \label{eq:energy}
\end{equation}
A low value of $E_{i,j}$ indicates that the dynamic state remains compatible with the original local content; a high value signals drift or incompatibility.

\noindent\textbf{Gradient-Based Forcing Term.}
Inspired by classical mechanics, where a force drives a system toward lower potential energy via $\mathbf{F} = -\nabla V$, we define a corrective force that acts on the dynamic hidden state to pull it toward energetically favorable configurations relative to the static reference:
\begin{equation}
    \mathbf{F}_{i,j} = -\eta\, \nabla_{\mathbf{h}_{i,j}} \mathcal{E}(\mathbf{h}_{i,j}, \boldsymbol{\phi}_{i,j})
    \label{eq:force}
\end{equation}
where $\eta$ is a learnable positive scalar controlling the overall force magnitude. The gradient $\nabla_{\mathbf{h}} \mathcal{E}$ is computed via the chain rule: since $\mathbf{z} = [\mathbf{h}^\top, \boldsymbol{\phi}^\top]^\top$, the Jacobian $\partial \mathbf{z}/\partial \mathbf{h} = [\mathbf{I}_C;\, \mathbf{0}]^\top$, so $\nabla_{\mathbf{h}} \mathcal{E}$ is simply the first $C$ components of the full gradient $\nabla_{\mathbf{z}} \mathcal{E}$, computed efficiently via automatic differentiation.

\begin{figure}[t]
    \centering
    \includegraphics[width=0.95\linewidth]{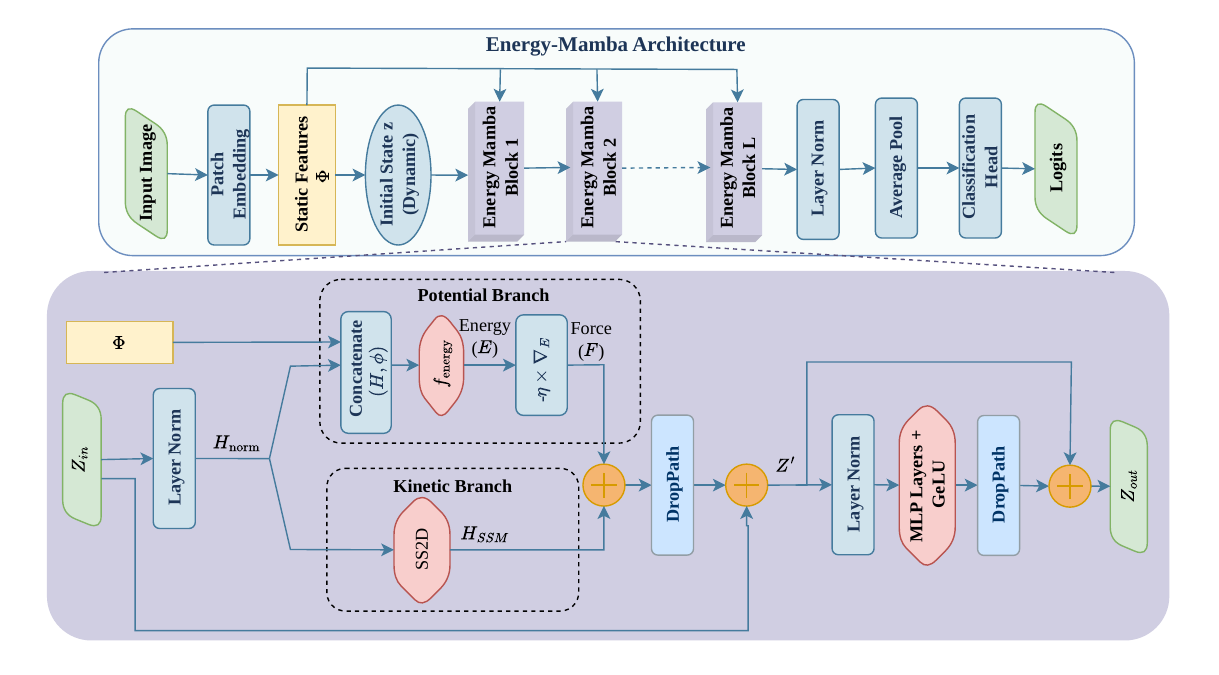}
    \caption{The Energy-Mamba block comprises two parallel branches: a kinetic branch (SS2D) capturing long-range spatiotemporal dependencies, and a potential energy branch deriving corrective forces from the gradient of a learned energy function. Their outputs are fused via a residual connection and further refined by a channel-wise MLP.}
    \label{fig:archt}
\end{figure}

\noindent\textbf{Block Architecture.}
As illustrated in Fig.~\ref{fig:archt}, the Energy-Mamba block consists of two parallel branches followed by a channel-wise MLP. The \textit{kinetic branch} applies SS2D to the normalized state to capture long-range spatial dependencies:
\begin{equation}
    \mathbf{H}_{\mathrm{ssm}} = \mathrm{SS2D}(\mathbf{H}_{\mathrm{norm}})
    \label{eq:kinetic}
\end{equation}
The \textit{potential branch} computes the spatial force map $\mathbf{F}$ over all locations using Eq.~\ref{eq:force}. Both contributions are fused via a residual connection with stochastic depth regularization~\cite{huang2016deep}:
\begin{equation}
    \mathbf{Z}' = \mathbf{Z} + \mathrm{DropPath}(\mathbf{H}_{\mathrm{ssm}} + \mathbf{F})
    \label{eq:combined}
\end{equation}
A final MLP refines channel-wise interactions:
\begin{equation}
    \mathbf{Z}_{\mathrm{out}} = \mathbf{Z}' + \mathrm{DropPath}\!\left(\mathrm{MLP}(\mathrm{LayerNorm}(\mathbf{Z}'))\right)
    \label{eq:mlp}
\end{equation}

\textbf{Hamiltonian Interpretation.}
The Energy-Mamba block admits an interpretation as a discretized Hamiltonian system with total energy $\mathcal{H} = T + V$. The SS2D module embodies the \textit{kinetic term} $T$, governing how state information propagates across space through learned SSM dynamics, analogous to the momentum-driven evolution of a physical system. The energy function $\mathcal{E}(\mathbf{H}_{\mathrm{norm}}, \mathbf{\Phi})$ serves as the \textit{potential term} $V$, defining a low-energy manifold of configurations compatible with the local image content. The forcing term $\mathbf{F} = -\eta\nabla\mathcal{E}$ implements the classical principle that forces drive systems toward lower potential, continuously pulling the dynamic state back toward this manifold when drift occurs. The combined update (Eq.~\eqref{eq:combined}) represents a single integration step of a system evolving under both internal dynamics (SSM) and an external potential field.

This contrasts with standard Mamba, which operates as a purely kinetic system with no restoring potential, analogous to a ball rolling on a flat surface, where any perturbation propagates freely without correction. Energy-Mamba instead resembles a ball rolling in a valley: the energy gradient continuously guides the trajectory toward stable, content-consistent representations, improving robustness to noise and input ordering effects. By jointly learning the \textit{rules of motion} (the SSM) and the \textit{rules of constraint} (the energy function), the block unifies interpretability and performance within a single coherent framework.

\section{Experiments \& Results}

\textbf{Datasets:}
To evaluate our proposed approach, we utilized four 2D datasets from MedMNIST~\cite{medmnistv1,medmnistv2}, a large-scale benchmark of standardized biomedical images spanning diverse modalities and classification tasks. Specifically, we selected OCTMNIST (retinal OCT), PneumoniaMNIST (chest X-ray), BloodMNIST (blood cell microscopy), and OrganCMNIST (abdominal CT), covering a representative range of imaging modalities, class cardinalities, and dataset scales, as summarized in Table~\ref{tab:datasets}. All images were resized to $224\times224$ and normalized prior to training. To improve generalization under limited data conditions, standard augmentation techniques were applied during training, including random resizing, cropping, horizontal flipping, and random erasing. The original train/validation/test splits defined in~\cite{medmnistv2} were preserved without modification.

\begin{table}[ht]
\centering
\renewcommand{\arraystretch}{1}
\resizebox{0.9\textwidth}{!}{%
\begin{tabular}{l l c l}
\hline
\textbf{Dataset} & \textbf{Modality} & \textbf{\# Classes} & \textbf{Train / Val / Test} \\
\hline
OCTMNIST      & Retinal OCT          & 4  & 97,477 / 10,832 / 1,000 \\
PneumoniaMNIST & Chest X-Ray         & 2  & 4,708 / 524 / 624       \\
BloodMNIST    & Blood Cell Microscopy & 8  & 11,959 / 1,712 / 3,421  \\
OrganCMNIST   & Abdominal CT         & 11 & 12,975 / 2,392 / 8,216  \\
\hline
\end{tabular}}
\caption{Summary of the four MedMNIST datasets used in this study.}
\label{tab:datasets}
\end{table}

\textbf{Benchmark Models:} 
We compare Energy-Mamba against seven representative architectures spanning three major paradigms in medical image classification. From the CNN family, we include \textit{ResNet-18} and \textit{ResNet-50}~\cite{he2016deep}, two widely adopted residual networks that serve as strong convolutional baselines, offering efficient local feature extraction but limited capacity for long-range spatial reasoning. From the Transformer family, we evaluate \textit{ViT-B}~\cite{dosovitskiy2020image}, a standard Vision Transformer that models global dependencies via self-attention at quadratic complexity, and \textit{Swin-T}~\cite{liu2021swin}, a hierarchical Transformer with shifted window attention that reduces this cost while preserving multi-scale representations. We also include \textit{MedViT}~\cite{manzari2023medvit}, a Transformer variant specifically designed for medical image analysis. From the SSM family, we evaluate \textit{MedMamba}~\cite{yue2024medmamba}, a hybrid CNN-SSM architecture that integrates convolutional local feature extraction with selective state-space scanning for medical classification, and \textit{MedKAFormer-T}~\cite{wang2025medkaformer}, a recent medical vision model combining Kolmogorov-Arnold Networks~\cite{liu2024kan} with Transformer-style attention.   

\textbf{Implementation Details:}
Unlike hierarchical architectures such as MedMamba~\cite{yue2024medmamba}, which employ a pyramidal design with patch-merging layers that progressively downsample spatial resolution while expanding channel depth, Energy-Mamba adopts a uniform, non-hierarchical structure. Following initial patch embedding, which downsamples the input to $28\times28$, both spatial resolution and channel dimension remain constant across all Energy-Mamba blocks. The potential energy MLP comprises three linear layers with two GeLU activations, while the channel refinement MLP uses two linear layers with one GeLU activation. Energy-Mamba is implemented in PyTorch and trained on a single NVIDIA RTX 3090 GPU. We used a batch size of 32, an initial learning rate of $1\times10^{-3}$, and a weight decay of $1\times10^{-4}$, optimized using the Adam optimizer with a cosine annealing learning rate scheduler. Models were trained for up to 50 epochs.

\textbf{Results:}
Table~\ref{tab:results} reports the AUC, accuracy, and F1 score of Energy-Mamba and all benchmark models across four MedMNIST datasets. Despite comprising only 2.0M parameters, roughly 6$\times$ fewer than the next most compact competitor (MedKAFormer-T at 12.47M) and over 43$\times$ fewer than ViT-B, Energy-Mamba achieves state-of-the-art or highly competitive performance across all evaluated datasets. On \textit{OCTMNIST} and \textit{BloodMNIST}, Energy-Mamba achieves the highest scores across all three metrics. For \textit{OCTMNIST}, it surpasses the second-best method, MedKAFormer-T, by margins of 1.45\%, 6.57\%, and 7.50\% in AUC, accuracy, and F1, respectively. On \textit{BloodMNIST}, it leads with an AUC of 99.91\%, accuracy of 97.79\%, and F1 of 97.68\%, outperforming all baselines including larger architectures such as ResNet-50 and Swin-T. On \textit{PneumoniaMNIST} and \textit{OrganCMNIST}, Energy-Mamba ranks first in AUC on both datasets and achieves the best F1 on \textit{OrganCMNIST}, with accuracy remaining competitive and within the top two across both benchmarks.

\begin{table*}[htbp]
\centering
\renewcommand{\arraystretch}{1.1}
\resizebox{\textwidth}{!}{%
\begin{tabular}{|l|c|c|c|c|c|c|c|c|}
\hline
\multirow{2}{*}{\textbf{Method}} & \multirow{2}{*}{\textbf{\#Params}}
  & \multirow{2}{*}{\textbf{\#FLOPs}}
  & \multicolumn{3}{c|}{\textbf{OCTMNIST}}
  & \multicolumn{3}{c|}{\textbf{PneumoniaMNIST}} \\
\cline{4-9}
& & & AUC (\%)$\uparrow$ & ACC (\%)$\uparrow$ & F1 (\%)$\uparrow$
  & AUC (\%)$\uparrow$ & ACC (\%)$\uparrow$ & F1 (\%)$\uparrow$ \\
\hline
ResNet-18       & 11.70 M & 1.82 G & 95.06 $\pm$ 0.83  & 77.20 $\pm$ 2.72 & 76.20 $\pm$ 1.41 & 96.01 $\pm$ 0.71 & 84.78 $\pm$ 1.81 & 81.82 $\pm$ 2.51\\
ResNet-50       & 25.60 M & 4.11 G & 95.84 $\pm$ 1.03 & 79.47 $\pm$ 0.06 & 77.95 $\pm$ 0.13 & 94.96 $\pm$ 0.49 & 83.81 $\pm$ 0.60 & 80.75 $\pm$ 0.41\\
ViT-B           & 86.57 M & 17.58 G & 90.89 $\pm$ 0.20 & 63.23 $\pm$ 0.90 & 56.28 $\pm$ 2.21 & 92.91 $\pm$ 1.12 & 79.91 $\pm$ 2.52 & 75.52 $\pm$ 3.94 \\
Swin-T          & 28.29 M & 4.50 G & 94.47 $\pm$ 0.36 & 73.23 $\pm$ 2.02 & 70.23 $\pm$ 2.81 & 91.08 $\pm$ 1.50 & 84.46 $\pm$ 2.21 & 81.62 $\pm$ 3.06 \\
MedViT          & 32.16 M & 6.55 G & 96.78 $\pm$ 0.56 & 77.60 $\pm$ 1.91  & 75.35 $\pm$ 2.02 & 95.97 $\pm$ 0.71 & 84.31 $\pm$ 0.41 & 81.12 $\pm$ 0.73\\
MedMamba        & 14.48 M & 6.71 G & 95.90 $\pm$ 0.55 & 77.37 $\pm$ 0.29 & 74.98 $\pm$ 0.12 & 94.89 $\pm$ 1.23 & 87.29 $\pm$ 3.69 & 85.16 $\pm$ 4.87\\
MedKAFormer-T   & 12.47 M & 7.80 G & \second{97.38 $\pm$ 0.43} & \second{81.03 $\pm$ 0.59 } & \second{80.00 $\pm$ 0.56} & \second{96.11 $\pm$ 0.43} & \second{90.43 $\pm$ 0.50 } & \best{89.99$\pm$ 1.13} \\
\hline
\textbf{Energy-Mamba} & \textbf{2.0 M} & \textbf{2.18 G} & \best{98.83 $\pm$ 0.15} & \best{87.60 $\pm$ 1.73} & \best{87.50 $\pm$ 1.82} & \best{97.04 $\pm$ 0.26 } & \best{90.74 $\pm$ 1.26} & \second{87.44 $\pm$ 1.05} \\
\hline
\end{tabular}}
\vspace{1pt}
\resizebox{\textwidth}{!}{%
\begin{tabular}{|l|c|c|c|c|c|c|c|c|}
\hline
\multirow{2}{*}{\textbf{Method}} & \multirow{2}{*}{\textbf{\#Params}}
  & \multirow{2}{*}{\textbf{\#FLOPs}}
  & \multicolumn{3}{c|}{\textbf{BloodMNIST}}
  & \multicolumn{3}{c|}{\textbf{OrganCMNIST}} \\
\cline{4-9}
& & & AUC (\%)$\uparrow$ & ACC (\%)$\uparrow$ & F1 (\%)$\uparrow$
  & AUC (\%)$\uparrow$ & ACC (\%)$\uparrow$ & F1 (\%)$\uparrow$ \\
\hline
ResNet-18       & 11.70 M & 1.82 G & \second{99.77 $\pm$ 0.02} & 96.35 $\pm$ 0.15 & 95.71 $\pm$ 0.29 & 98.21 $\pm$ 0.41 & 88.22 $\pm$ 2.97 & 86.24 $\pm$ 3.35 \\
ResNet-50       & 25.60 M & 4.11 G & 99.77 $\pm$ 0.03 & 96.04 $\pm$ 0.24 & 95.20 $\pm$ 0.20 & 98.77 $\pm$ 0.10 & 88.59 $\pm$ 0.17 & 87.03 $\pm$ 0.35 \\
ViT-B           & 86.57 M & 17.58 G & 98.55 $\pm$ 0.78 & 88.01 $\pm$ 5.54 & 85.10 $\pm$ 7.43 & 98.03 $\pm$ 0.17 & 83.38 $\pm$ 1.08 & 81.55 $\pm$ 1.03 \\
Swin-T          & 28.29 M & 4.50 G & 99.69 $\pm$ 0.02 & 96.22 $\pm$ 0.22 & 95.76 $\pm$ 0.27 & 98.29 $\pm$ 0.06 & 89.22 $\pm$ 0.12 & 88.08 $\pm$ 0.23 \\
MedViT          & 32.16 M & 6.55 G & 99.65 $\pm$ 0.04 & 94.59 $\pm$ 0.43 & 94.31 $\pm$  0.66 & 98.35 $\pm$ 0.20 & 90.66 $\pm$ 1.95 & 88.72 $\pm$ 0.62\\
MedMamba        & 14.48 M & 6.71 G & 99.15 $\pm$ 0.57 & 95.30 $\pm$ 1.00 & \second{96.68 $\pm$ 0.53} & {98.81 $\pm$ 0.14} & \best{91.48 $\pm$ 1.09} & \second{89.51 $\pm$ 0.51} \\
MedKAFormer-T   & 12.47 M & 7.80 G & {99.77 $\pm$ 0.12} & \second{96.52 $\pm$ 0.28} & 94.34 $\pm$ 0.59 & \second{99.10} $\pm$ 0.06 & 89.94 $\pm$ 1.02 & 88.91 $\pm$ 0.89 \\
\hline
\textbf{Energy-Mamba} & \textbf{2.0 M} & \textbf{2.18 G} & \best{99.91 $\pm$ 0.02} & \best{97.79 $\pm$ 0.31} & \best{97.68 $\pm$ 0.40} & \best{99.21 $\pm$ 0.03} & \second{90.80 $\pm$ 0.56 } & \best{90.18 $\pm$ 0.35} \\
\hline
\end{tabular}}
\caption{Comparison of AUC (\%), accuracy (ACC, \%), and F1 score (\%) across four MedMNIST datasets. \textcolor{red}{{Red}} denotes the best result and \textcolor{blue}{blue} the second best. }
\label{tab:results}
\end{table*}

\begin{figure}
    \centering
    \includegraphics[width=0.7\linewidth]{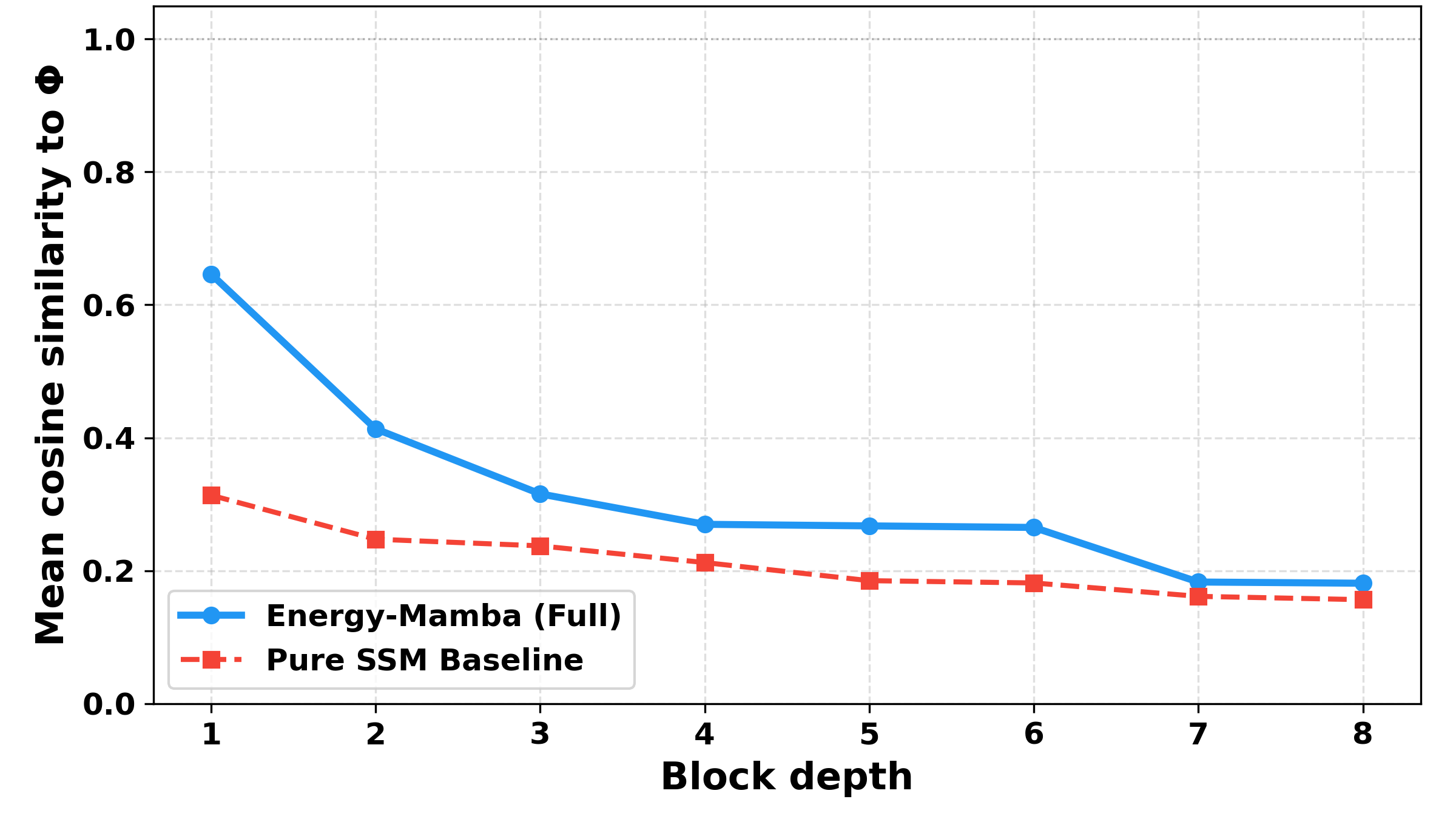}
    \caption{Mean cosine similarity between the static input representation and intermediate features at each block depth, comparing Energy-Mamba against a pure SSM baseline.}
    \label{fig:drift_representation}
\end{figure}

Crucially, these gains are achieved at a computational cost of just 2.18G FLOPs underscoring that the performance improvements do not come at the expense of efficiency. Heavier models such as ViT-B, despite requiring 17.58G FLOPs, consistently underperform across all datasets, suggesting that raw model capacity is not the primary driver of accuracy in this domain. Taken together, these results demonstrate that Energy-Mamba's design effectively prioritizes meaningful representational capacity over scale, yielding strong classification performance while remaining suitable for deployment in computationally constrained clinical settings.

To assess whether Energy-Mamba preserves initial feature structure throughout its depth, Figure~\ref{fig:drift_representation} tracks the mean cosine similarity between the static input representation $\Phi$ and intermediate features after each block, comparing the full architecture against a pure SSM baseline. Energy-Mamba maintains substantially higher similarity in early blocks (0.65 at block 1 vs. 0.31 for the baseline), indicating more faithful retention of the original representational content in shallow layers. Both models gradually converge to comparable values near 0.18–0.19 by block 8, which is expected as deeper blocks are responsible for constructing increasingly abstract representations. Nevertheless, the consistently higher similarity maintained by Energy-Mamba across intermediate blocks confirms that it achieves a more stable feature transformation across depth, largely addressing the representational drift present in standard SSM-based architectures while still permitting the necessary abstraction in later stages.

\section{Conclusion}

We presented Energy-Mamba, a physics-informed state-space model that addresses representational drift in vision Mamba architectures through a learnable potential energy function and a gradient-based forcing term. By grounding the hidden state dynamics in a Hamiltonian framework, the model jointly learns the rules of spatial propagation and the rules of representational constraint, enabling the dynamic state to remain faithful to local image content throughout the network. Evaluated across four MedMNIST datasets, Energy-Mamba achieves state-of-the-art or competitive performance on all benchmarks while comprising only 2.0M parameters, a fraction of the capacity required by competing approaches. These results suggest that explicitly constraining hidden state trajectories through physics-inspired potentials is a promising and parameter-efficient alternative to increasing model capacity or architectural complexity. Several directions remain open for future investigation, including the extension of Energy-Mamba to 3D volumetric medical data, the extension to other tasks such as segmentation, the exploration of alternative energy function architectures, and a systematic ablation of the interplay between the kinetic and potential branches across different imaging modalities and dataset scales.

\section*{Acknowledgments}
This work was supported under the International Research Collaboration Co-Fund (IRCC) between Qatar University and University of Hyogo, grant number IRCC-2025-633.

\bibliographystyle{splncs04}
\bibliography{refs}
\end{document}